\documentclass[twocolumn,twoside,slac_two]{revtex4}
\usepackage{graphicx}
\usepackage{fancyhdr}
\begin{document}

\title{A strong radio brightening at the jet base of M87 in the period of the
elevated TeV $\gamma$-ray state in 2012}

%

\author{K.~Hada$^{a,b}$, 
M.~Giroletti$^a$, 
M.~Kino$^{c,d}$, 
G.~Giovannini$^{a,e}$, 
F.~D'Ammando$^a$,
C.~C.~Cheung$^f$,
M.~Beilicke$^g$, 
H.~Nagai$^h$, 
A.~Doi$^d$,
K.~Akiyama$^{b,i}$, 
M.~Honma$^{b,j}$,
K.~Niinuma$^{k}$, 
C.~Casadio$^{l}$,
M.~Orienti$^a$, 
H.~Krawczynski$^7$, 
J.~L.~G\'omez$^{l}$, 
S.~Sawada-Satoh$^b$, 
S.~Koyama$^{b,d,i}$, 
A.~Cesarini$^{m}$, 
S.~Nakahara$^{n}$ and 
M.~A.~Gurwell$^{o}$
}

\affiliation{
$^a$INAF Istituto di Radioastronomia, via Gobetti 101, I-40129 Bologna,
Italy \\
$^b$Mizusawa VLBI Observatory, National Astronomical Observatory of Japan,
Osawa, Mitaka, Tokyo 181-8588, Japan \\
$^c$Korea Astronomy and Space Science Institute, 776 Daedukdae-ro, Yusong,
Daejon 305-348, Korea \\
$^d$Institute of Space and Astronautical Science, Japan Aerospace
Exploration Agency, 3-1-1 Yoshinodai, Chuo, Sagamihara 252-5210, Japan \\
$^e$Dipartimento di Fisica e Astronomia, Universit\`a di Bologna, via
Ranzani 1, I-40127 Bologna, Italy \\
$^f$Space Science Division, Naval Research Laboratory, Washington, DC
20375, USA \\
$^g$Physics Department and McDonnell Center for the Space Sciences,
Washington University, St. Louis, MO 63130, USA \\
$^h$National Astronomical Observatory of Japan, Osawa, Mitaka, Tokyo
181-8588, Japan \\
$^i$Department of Astronomy, Graduate School of Science, The University of
Tokyo, 7-3-1 Hongo, Bunkyo-ku, Tokyo 113-0033, Japan \\
$^{j}$Department of Astronomical Science, The Graduate University for
Advanced Studies (SOKENDAI), 2-21-1 Osawa, Mitaka, Tokyo 181-8588, Japan \\
$^{k}$Graduate School of Science and Engineering, Yamaguchi University, 1677-1
Yoshida, Yamaguchi, 753-8512, Japan \\
$^{l}$Instituto de Astrofisica de Andalucia, CSIC, Apartado 3004, 18080
Granada, Spain \\
$^{m}$Department of Physics, University of Trento, I38050, Povo, Trento,
Italy \\
$^{n}$Faculty of Science, Kagoshima University, 1-21-35 Korimoto, Kagoshima,
Kagoshima 890-0065, Japan \\
$^{o}$Harvard-Smithsonian Center for Astrophysics, Cambridge MA 02138 USA
}

\begin{abstract}
The nearby radio galaxy M87 offers a unique opportunity for exploring the
connection between $\gamma$-ray production and jet formation at an unprecedented
linear resolution. However, the origin and location of the $\gamma$-rays in this
source is still elusive. Based on previous radio/TeV correlation events, the
unresolved jet base (radio core) and the peculiar knot HST-1 at $>$120~pc from the
nucleus are proposed as candidate site(s) of $\gamma$-ray production. Here we
report our intensive, high-resolution radio monitoring observations of the M87 jet
with the VLBI Exploration of Radio Astrometry (VERA) and the European VLBI Network
(EVN) from February 2011 to October 2012, together with contemporaneous
high-energy $\gamma$-ray light curves obtained by the Fermi Large Area
Telescope. During this period, an elevated level of the M87 flux is reported at
TeV with VERITAS. We detected a remarkable flux increase in the radio core with
VERA at 22/43 GHz coincident with the VHE activity. Meanwhile, HST-1 remained
quiescent in terms of its flux density and structure in the radio band. These
results strongly suggest that the TeV $\gamma$-ray activity in 2012 originates in
the jet base within 0.03~pc (projected) from the central supermassive black hole.
\end{abstract}

\maketitle

\thispagestyle{fancy}


\section{Introduction}
The nearby radio galaxy M87 accompanies one of the best studied AGN jets. Its
proximity (16.7~Mpc) and brightness have enabled detailed studies of this jet over
decades through radio, optical and to X-ray at tens of parsec scale
resolutions. Furthermore, the inferred very massive black hole~($M_{\rm BH} \simeq
(3-6) \times 10^9~M_{\odot}$) yields a linear resolution down to $1~{\rm
milliarcsecond~(mas)} = 0.08~{\rm pc}=140$ Schwarzschild radii~$(R_{\rm s})$ (for
$M_{\rm BH} = 6 \times 10^9~M_{\odot}$), making this source an ideal case to probe
the relativistic-jet formation at an unprecedented compact scale with
Very-Long-Baseline-Interferometer (VLBI) observations~(e.g., Ly et al. 2007;
Kovalev et al. 2007; Hada et al. 2011; Asada \& Nakamura 2012; Doeleman et
al. 2012; Hada et al. 2013). M87 is now widely known to show $\gamma$-ray emission
up to the very-high-energy (VHE; $E>100$~GeV) regime, where this source often
exhibits active flaring episodes.  The location and the physical processes of such
emission have been a matter of debate over the past years, and there are two
candidate sites which can be responsible for the VHE $\gamma$-ray production. One
is a very active knot HST-1 which is located at more than 100 pc from the
nucleus~(Stawarz et al.  2006; Cheung et al. 2007; Harris et al. 2009). This
argument is based on the famous VHE flare event in 2005, where HST-1 underwent a
large radio-to-X-ray outburst jointly with a VHE flare. In contrast, the other
candidate is the core/jet base, which is very close to the central black
hole. This argument is based on the VHE event in 2008, where the core/VHE showed a
remarkable correlation in the light curves. There was another VHE event in 2010,
but this is rather elusive. Coincident with the VHE event, Chandra detected an
enhanced flux from the X-ray core~(Harris et al. 2011; Abramowski et al. 2012),
and VLBA observations also suggested a possible increase of the radio core
flux~(Hada et al. 2012). However, Giroletti et al. (2012) found the emergence of a
superluminal component in the HST-1 complex near the epoch of this event, which is
reminiscent of the 2005 case.

Recently, the VERITAS Collaboration has reported new VHE $\gamma$-ray activity 
from M87 in early 2012~(Beilicke et al. 2012). While there were no remarkable 
flares like those in the previous episodes, the VHE flux in 2012 clearly 
exhibits an elevated state at a level of $\sim$$9\sigma$ ($\Phi_{\rm > 0.35TeV}
\sim$(0.2--0.3)$\times 10^{-11}$ photons~cm$^{-2}$~s$^{-1}$) over the 
consecutive two months from February to March 2012. The observed flux is a 
factor of $\sim$2 brighter than that in the neighboring quiescent periods.  
Therefore, this event provides another good opportunity for exploring the 
location of the VHE emission site by jointly using high-resolution instruments. 

\section{Observations}
Here we report a multi-wavelength radio and MeV/GeV study of the M87 jet during
this period using the VLBI Exploration Radio Astrometry (VERA, Fegure 1), the
European VLBI Network (EVN), the Submillimeter Array (SMA) and the Fermi-LAT. We
especially focus on the VLBI data in the radio bands; with VERA, we obtained the
high-angular-resolution, dense-sampling-interval, phase-referencing data set at 22
and 43~GHz during the VHE activity in 2012: with the supportive EVN monitoring, we
obtained a complementary data set at 5~GHz, which enables a high-sensitivity
imaging of the M87 jet. A collective set of these radio data allows us to probe
the detailed physical status and structural evolutions of M87 by pinpointing the
candidate sites of the $\gamma$-ray emission i.e., the core and HST-1. For more
details regarding the radio data analysis, see Hada et al.~(2014).

The LAT data reported here were collected from 2011 February 1 (MJD 55593) to 2012
September 30 (MJD 56200). During this time, the Fermi observatory operated almost
entirely in survey mode. The analysis was performed with the \texttt{ScienceTools}
software package version v9r32p5.  The LAT data were extracted within a
$10^{\circ}$ region of interest centred at the radio location of M87. Only events
belonging to the `Source' class were used. The time intervals when the rocking
angle of the LAT was greater than 52$^{\circ}$ were rejected. In addition, a cut
on the zenith angle ($< 100^{\circ}$) was applied to reduce contamination from the
Earth limb $\gamma$ rays, which are produced by cosmic rays interacting with the
upper atmosphere.  The spectral analysis was performed with the instrument
response functions \texttt{P7REP\_SOURCE\_V15} using an unbinned
maximum-likelihood method implemented in the Science tool \texttt{gtlike}. A
Galactic diffuse emission model and isotropic component, which is the sum of an
extragalactic and residual cosmic ray background, were used to model the
background.  The normalizations of both components in the background model were
allowed to vary freely during the spectral fitting.

We evaluated the significance of the $\gamma$-ray signal from the sources by means
of the maximum-likelihood test statistic TS = 2$\Delta$log(likelihood) between
models with and without a point source at the position of M87 (Mattox et
al. 1996). The source model used in \texttt{gtlike} includes all of the point
sources from the second Fermi-LAT catalog (2FGL; Nolan et al. 2012) that fall
within $15^{\circ}$ of the source. The spectra of these sources were parametrized
by power-law functions, except for 2FGL\,J1224.9+2122 (4C\,21.35) and
2FGL\,J1229.1+0202 (3C\,273), for which we used a log-parabola as in the 2FGL
catalogue. A first maximum-likelihood analysis was performed to remove from the
model the sources having TS $<$ 25 and/or the predicted number of counts based on
the fitted model $N_{pred} < 3 $. A second maximum-likelihood analysis was
performed on the updated source model. In the fitting procedure, the normalization
factors and the photon indices of the sources lying within 10$^{\circ}$ of M87
were left as free parameters. For the sources located between 10$^{\circ}$ and
15$^{\circ}$, we kept the normalization and the photon index fixed to the values
from the 2FGL catalogue.

\begin{figure}[ttt]
\centering \includegraphics[width=\columnwidth]{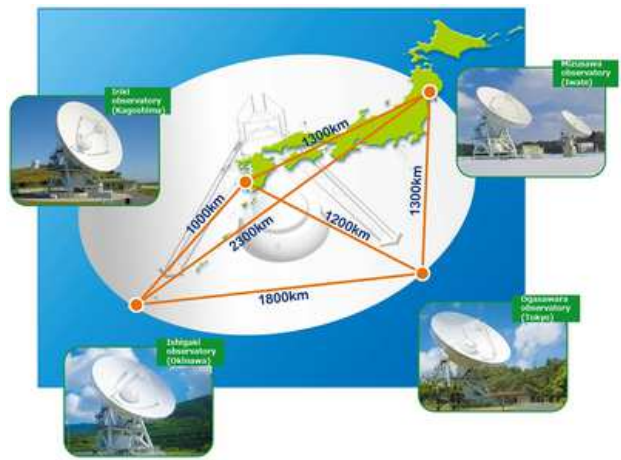} \caption{Layout of the
VLBI Exploration of Radio Astrometry (VERA).}
\end{figure}

\begin{figure*}[htbp]
\centering
\includegraphics[width=0.9\textwidth]{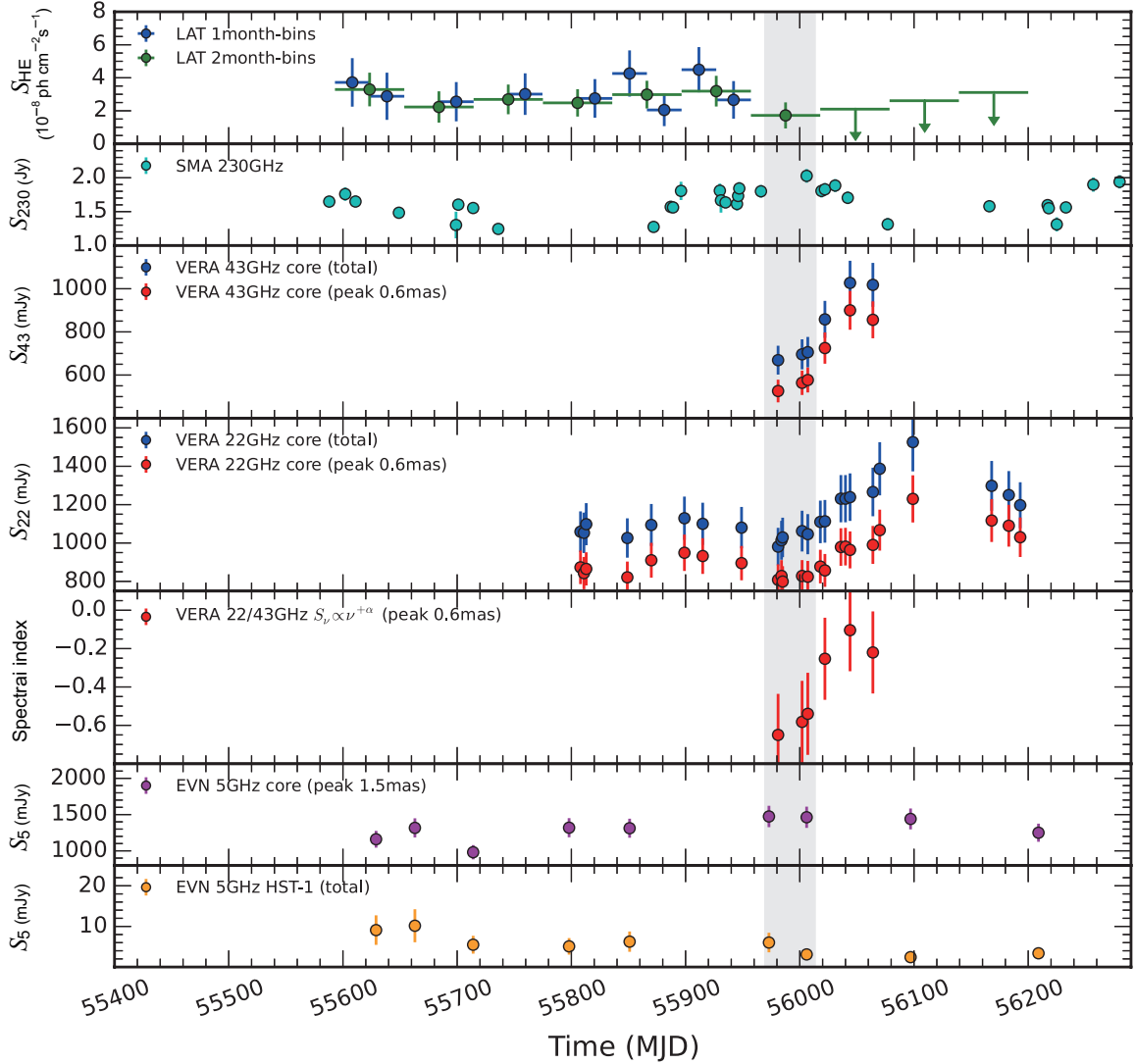}
\caption{Multi-wavelength light curves of M87 between 2011 February and
2012 December. The vertical shaded area over the plots indicates a period of
elevated VHE emission reported by Beilicke et al. (2012).}
\end{figure*}

Integrating over the period from 2011 February 1 to 2012 September 30 (MJD
55593--56200), the fit with a power-law model in the 0.1--100 GeV energy range
results in a TS = 134, with an integrated average flux of (2.22 $\pm$ 0.43)
$\times$10$^{-8}$ ph cm$^{-2}$ s$^{-1}$ and a photon index of $\Gamma$ = 2.25
$\pm$ 0.10. Taking into account the detection significance over the whole analysed
period, we produced the $\gamma$-ray light curves with 1-month and 2-month time
bins. This choice of binning is compatible with those adopted in the previous M87
studies with LAT data~(Abdo et al. 2009; Abramowski et al. 2012), and also
reasonable for a comparison with the observed month-scale VHE activity in
2012. For each time bin, the spectral parameters for M87 and for all the sources
within 10$^{\circ}$ from it were frozen to the value resulting from the likelihood
analysis over the entire period. In the light curve with the 2-month time bins, if
TS $<$ 10, 2$\sigma$ upper limits were evaluated, while only bins with TS $>$ 10
are selected in the light curve with the 1-month time bins. We describe the
results of the LAT light curves in Section~4.2.

Dividing the 1-month bins with higher flux in 5-day sub-bins, the highest flux of
(10.4$\pm$4.8)$\times$10$^{-8}$ and (8.6$\pm$3.4)$\times$10$^{-8}$ ph cm$^{-2}$
s$^{-1}$ was detected on 2011 October 12-16 and 2012 January 16-20, respectively
(these sub-bin data also show TS$>$10). By means of the \texttt{gtsrcprob} tool,
we estimated that the highest energy photon emitted from M87 (with probability
$>$ 90\% of being associated with the source) was observed by LAT on 2011 April 7,
at a distance of 0.09$^{\circ}$ from the source and with an energy of 254.0 GeV,
extending into the VHE range.

\begin{figure*}[ttt]
\centering \includegraphics[width=0.9\textwidth]{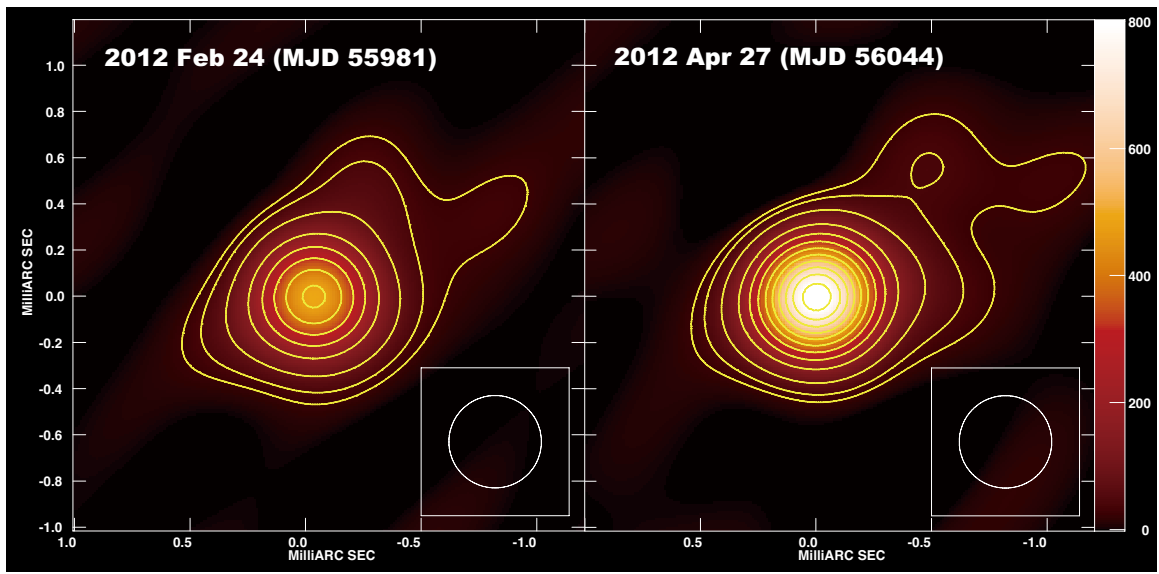} \caption{VERA 43
GHz images of the M87 jet during the elevated VHE state in 2012.} \end{figure*}

\subsection{Results}
In Figure~2 we show a combined set of light curves of M87 from radio to MeV/GeV
$\gamma$-ray between MJD~55400 and MJD~56280. Thanks to the dense, complementary
coverages of VERA and EVN, we revealed the detailed evolutions of the radio light
curves for both the core and HST-1. The most remarkable finding in these plots is
a strong enhancement of the radio core flux at VERA 22 and 43~GHz, which occurred
coincidentally with the elevated VHE state. At 22~GHz, we further detected a
subsequent decay stage of the brightness at the last three epochs. Also at 43~GHz,
we detected possible saturation of the flux increase near the last
epoch. Meanwhile, the EVN monitoring confirmed a constant decrease of the HST-1
luminosity.  Figure~3 describes VERA 43-GHz images during the VHE active period,
which indicate the flux enhancement within the central resolution element of
0.4~mas, corresponding to a linear scale of 0.03 pc or 56~$R_{\rm s}$. We also
note that the SMA data at 230~GHz also appear to show a local maximum in its light
curve during the period of the elevated VHE state.

Another notable finding is a frequency-dependent evolution of the radio core
flare. The VERA light curves clearly indicate that the radio core brightens more
rapidly with a larger amplitude as frequency increases. At 43~GHz, the flux
increased up to $\sim$70\% for the subsequent 2 months at an averaged rate of
$\sim$35\%/month, and afterward the growth seems to be saturated. On the other
hand, the core flux at 22-GHz progressively increased up to $\sim$50\% for the
subsequent 4 months at a slower rate of $\sim$12\%/month. At 5~GHz, by contrast,
the core remained virtually stable within the adopted error of 10\%. This is the
first time that such a frequency-dependent nature of the radio flare is clearly
confirmed in the M87 jet. We also detected a core-shift between 22 and 43~GHz by
using the VERA dual-beam astrometry technique (see Hada et al. 2014), where the
amount of the shift was similar to the value obtained in the previous core-shift
measurement (Hada et al. 2011).

Regarding the MeV/GeV regime, the LAT light curves were stable up to February
2012, and we did not find any significant flux enhancement during the period of
the VHE activity.  After March 2012, however, no significant emission was detected
for the subsequent 6 months in the 1- and 2-month binned data, suggesting a change
in the HE state after the VHE event.  This indicates a decrease in the HE flux (by
a factor of $\sim 2$) after the VHE event, in agreement with the level of decrease
observed at VHE in 2012 April-May (Beilicke et al. 2012).

\section{Discussion and summary}
Following the 2008 episode this is the second time where a VHE event
accompanied a remarkable radio flare from the core. Meanwhile, the radio
luminosity of the HST-1 region was continuously decreasing, and we did not find
any hints of the emergence of new components from HST-1 as seen in 2005 and
2010. These results strongly suggest that the VHE activity in 2012 is associated
with the core at the jet base, while HST-1 is an unlikely source. We note
that these remarkable flares are very rare also in radio bands~(Acciari et al. 2009), 
so it is unlikely that an observed joint
radio/VHE correlation is a chance coincidence, while the low statistics of the LAT
light curves still do not allow conclusive results on the HE/VHE
connection. 

What kinds of mechanisms are responsible for the VHE production in the M87 core?
Some of the existing models ascribe the VHE production to extremely compact
regions near the central black hole (e.g., Neronov \& Aharonian 2007; Lenain et
al. 2008; Giannios et al. 2010; Barkov et al. 2012).  These models well explain
the rapid (a few days) variability observed in the previous VHE flares in 2005,
2008 and 2010.  However as far as we consider the case in 2012, the size of the
associated region expected from these models seems to be smaller than that
suggested by VLBI and the observed longer timescale of the VHE
variability. Indeed, a contemporaneous mm-VLBI observation at 230~GHz during the
2012 event also suggests the possible extended nature for the flaring region
($\gtrsim$0.3~mas; Akiyama et al. submitted).

Another popular scenario for the M87 VHE production comes from a blazar-type,
two-zone emission model where the VHE emission originates in the upstream part of
a decelerating jet (Georganopoulos et al. 2005) or in the layer part of the
spine-sheath structure (Tavecchio \& Ghisellini 2008).  However in their steady
state models, whether the models can explain the observed simultaneous radio/VHE
correlation or not has not been well investigated yet because the emission regions
associated with radio and VHE are spatially separated from each other. In this
respect, a simple, homogeneous one-zone synchrotron self-Compton jet model
examined by Abdo et al. (2009) would be interesting to note since one can in
principle accept coincident radio/VHE correlations.

Our multi-frequency radio monitoring additionally revealed a frequency-dependent
evolution of the radio light curves for the M87 core. Such a behavior is often
explained by the creation of a plasma condensation, which subsequently expands and
propagates down the jet under the effect of synchrotron-self-absorption (SSA). The
stronger SSA opacity at the jet base causes a delayed brightening at lower
frequencies, and the light curve at each frequency reaches its maximum when the
newborn component passes through the $\tau_{\rm ssa}(\nu)\sim 1$ surface (i.e.,
the radio core at the corresponding frequency). In this context, by jointly using
the observed time-lag ($\Delta t_{\rm 43-22}$) and core-shift ($\Delta r_{\rm
proj, 43-22}$), we can estimate an apparent speed of the propagating component
such that $\beta_{\rm app, 43 \rightarrow 22} = \frac{\Delta r_{\rm proj,
43-22}}{c \Delta t_{\rm 43-22}}$. This results in a speed about
$\sim$0.04$c$--0.22$c$, suggesting that the newborn component is sub-relativistic.
This is significantly smaller than the super-luminal features appeared from the
core during the previous VHE event in 2008~(1.1$c$; Acciari et al. 2009), where
the peak VHE flux is $\gtrsim$5 times higher than that in 2012. If we assume that
propagating shocks or component motions seen in radio observations reflect the
bulk velocity flow, this may suggest that the stronger VHE activity is associated
with the production of the higher Lorentz factor jet.

We are currently upgrading our M87 monitoring project by using the KVN and VERA
Array (KaVA; Niinuma et al. 2014), which dramatically improves jet imaging
capability thanks to the increase of the number of telescopes/baselines plus the
addition of shorter baselines. This will enable us to constrain the jet kinematics
and radio/VHE connection more precisely.

\bigskip 
\begin{acknowledgments}
The VERA is operated by Mizusawa VLBI Observatory, a branch of National
Astronomical Observatory of Japan.  e-VLBI research infrastructure in Europe is
supported by the European Union's Seventh Framework Programme (FP7/2007-2013)
under grant agreement no. RI-261525 NEXPReS. The European VLBI Network is a joint
facility of European, Chinese, South African and other radio astronomy institutes
funded by their national research councils.  The Submillimeter Array is a joint
project between the Smithsonian Astrophysical Observatory and the Academia Sinica
Institute of Astronomy and Astrophysics and is funded by the Smithsonian
Institution and the Academia Sinica.

The Fermi LAT Collaboration acknowledges generous ongoing support from a number of
agencies and institutes that have supported both the development and the operation
of the LAT as well as scientific data analysis. These include the National
Aeronautics and Space Administration and the Department of Energy in the United
States, the Commissariata l'Energie Atomique and the Centre National de la
Recherche Scientifique Institut National de Physique Nucl\'eaire et de Physique
des Particules in France, the Agenzia Spaziale Italiana and the Istituto Nazionale
di Fisica Nucleare in Italy, the Ministry of Education, Culture, Sports, Science
and Technology (MEXT), High Energy Accelerator Research Organization (KEK) and
Japan Aerospace Exploration Agency (JAXA) in Japan, and the K. A.Wallenberg
Foundation, the Swedish Research Council and the Swedish National Space Board in
Sweden. Additional support for science analysis during the operations phase is
gratefully acknowledged from the Istituto Nazionale di Astrofisica in Italy and
the Centre National dEtudes Spatiales in France.
\end{acknowledgments}

\bigskip 

\end{document}